\documentstyle[aps,prl,epsf,twocolumn,floats]{revtex}

\begin{document}
\draft
\title{Density of states in coupled chains with
off-diagonal disorder}

\author{P. W. Brouwer,$^a$ C. Mudry,$^a$ and A.\ Furusaki$^b$}
\address{$^a$Lyman Laboratory of Physics, Harvard University, Cambridge MA 02138\\$^b$Yukawa Institute for Theoretical Physics, Kyoto University,
Kyoto 606-8502, Japan\\
{\rm \today, 1999}
\medskip \\ \parbox{14cm}{\rm We compute the density of states
$\rho(\varepsilon)$ in $N$ coupled chains with random hopping. At zero
energy, $\rho(\varepsilon)$ shows a singularity that strongly depends
on the parity of $N$. For odd $N$, $\rho(\varepsilon) \propto
1/|\varepsilon \ln^3 \varepsilon|$, with and without time-reversal
symmetry. For even $N$, $\rho(\varepsilon) \propto |\ln \varepsilon|$ in
the presence of time-reversal symmetry, while there is a pseudogap,
$\rho(\varepsilon) \propto |\varepsilon \ln \varepsilon|$, in the
absence of time-reversal symmetry. \smallskip \\
PACS numbers: 71.55.Jv, 71.23.-k, 72.15.Rn, 11.30.Rd}}
\narrowtext
\maketitle 


Since the works of Dyson, it has been known that one dimensional
systems with off-diagonal disorder may show anomalous behavior in their
density of states and their localization properties \cite{Dyson1953}.
This anomalous behavior has reappeared in many reincarnations, ranging
from Dyson's original application of masses coupled with springs, to
the random hopping model \cite{TheodorouCohen,EggarterRiedinger},
quantum XY spin chains \cite{Fisher}, supersymmetric quantum mechanics
\cite{Comtet}, one-dimensional classical diffusion in a random medium
\cite{Bouchaud}, narrow-gap semiconductors \cite{OE}, and charge density
wave materials \cite{Lee}. In all these systems, it has been shown that
both the density of states and the localization length diverge at zero
energy. These two divergencies are closely linked by a theorem by
Thouless \cite{Thouless}.

Recently, it was found that the localization length of a wire
consisting of $N$ of such chains shows a remarkable dependence on the
parity of $N$ \cite{MillerWang,BMSA}: At zero energy, i.e.\ at the
center of the band, the localization length diverges for odd $N$
(odd $N$ includes the pure one-dimensional case $N=1$), while wavefunctions
remain localized for even $N$. Moreover, it was found that staggering
in the hopping parameter (as the consequence of a Peierls instability),
may lead to an additional set of delocalized states for all $N > 1$,
regardless of parity \cite{BMSA}. For comparison, in the pure
one-dimensional case $N=1$ staggering always enhances localization.  In
this paper we consider the density of states (d.o.s.) for the $N$-chain
wire with off-diagonal disorder.  Unlike in the one-dimensional case,
this is a problem that requires separate attention, since the Thouless
theorem which links d.o.s.\ and localization length does not hold for
$N>1$ coupled chains. Moreover this problem is also of relevance for
the random flux model, which is a special example of off-diagonal
randomness, and whose localization properties and density of states
near $\varepsilon=0$ are the subject of an ongoing debate
\cite{RandomFlux}.

To be specific, we consider the d.o.s.\ for the Schr\"odinger
equation
\begin{equation}
  \varepsilon \psi_n = 
  {\cal H} \psi_n = -t_{n}^{\vphantom{\dagger}} \psi_{n+1}
   - t_{n-1}^{\dagger} \psi_{n-1}, \label{eq:Schrod}
\end{equation}
where $\psi_n$ is an $N$-component wavefunction and $t_n$ an $N \times
N$ hopping matrix. The index $n=1,\ldots,L$ labels the site index along
the wire. The length of the wire, measured in units of the lattice
spacing, is $L$.  The reason why this system can display anomalous
behavior near zero energy is the
existence of a particle-hole or {\em chiral} symmetry, which is absent
in the presence of on-site disorder: Under a mapping $\psi_n \to (-1)^n
\psi_n$, the Hamiltonian ${\cal H}$ changes sign, ${\cal H} \to - {\cal
H}$. As a result, the eigenvalues of ${\cal H}$ occur in pairs $\pm
\varepsilon$. There are two mechanisms by which the chiral symmetry is
known to affect the d.o.s.\ near $\varepsilon=0$. First, level
repulsion of the eigenvalue $\varepsilon$ with its mirror image
$-\varepsilon$ causes a universal suppression of the d.o.s.\ near
$\varepsilon=0$ \cite{RMT}. This suppression appears on the scale of a
level spacing, is independent of the geometry, but becomes unimportant
in the thermodynamic limit. The second mechanism, for which several
different descriptions exist
\cite{TheodorouCohen,EggarterRiedinger,Fisher,Comtet,OE}, is special
for a (quasi) one-dimensional geometry, and survives in the
thermodynamic limit.  It is responsible for the divergence of the
d.o.s.\ in the pure one-dimensional case. Below we present a
calculation of the d.o.s.\ in the multi-chain case, combining
ideas from the
Fokker-Planck approach to localization in multichannel quantum wires
\cite{Transport} and the calculation of the d.o.s.\ in the
one-dimensional random hopping model \cite{EggarterRiedinger,OE}.

We first state our main results. We find that the parity dependence
that was previously obtained for the localization properties is also
present in the density of states. For odd $N$, the d.o.s.\ diverges at
zero energy according to
\begin{equation}
  \rho(\varepsilon) \propto {w^2 \over |\varepsilon\, \ln^3(w^2 /\varepsilon)|},\ \ \mbox{$N$ odd}, \label{eq:N1div} \label{eq:N1}
\end{equation}
where $w$ is a dimensionless parameter governing the randomness in the
$t_n$. The bandwidth is chosen as the unit of energy. In
Eq.\ (\ref{eq:N1}) and in the remainder of the paper we assume
$\varepsilon > 0$. The d.o.s.\ for $\varepsilon < 0$ then follows from
$\rho(-\varepsilon) = \rho(\varepsilon)$. The form of the divergence
(\ref{eq:N1div}) is independent of whether time-reversal symmetry is
broken or not, i.e.\ of whether the hopping matrices $t_n$ are
generically real or complex. For even $N$, in contrast, the density of
states strongly depends on the presence or absence of time-reversal
symmetry (labeled by the parameter $\beta = 1$ or $2$, respectively),
\begin{equation}
  \rho(\varepsilon) \propto 
  |\varepsilon/w^2|^{\beta-1} |\ln(w^2/\varepsilon)|,\ 
  \ \mbox{$N$ even}. \label{eq:N2div} \label{eq:two}
\end{equation}
In the presence of time-reversal symmetry, $\rho(\varepsilon)$ shows a
logarithmic divergence as $\varepsilon \to 0$, while in the absence of
time-reversal symmetry a pseudo-gap is opened, $\rho(\varepsilon)$
vanishes at $\varepsilon=0$. Such a strong dependence on time-reversal
symmetry is a remarkable result. An effect of comparable magnitude
appears in the suppression of the gap by a weak magnetic field in a
normal metal in the proximity of a superconductor. In the remainder of
this paper we derive the results (\ref{eq:N1div}) and (\ref{eq:N2div})
assuming a specific statistical model for the hopping matrices $t_n$ in
Eq.\ (\ref{eq:Schrod}). The details of this model are not relevant, the
singular behavior of the d.o.s.\ near zero energy being governed by the
fundamental symmetries of the Hamiltonian ${\cal H}$ only. As an
illustration of the general validity of our result, we close with a
comparison to numerical simulations.

As we are interested in the d.o.s.\ of the random hopping
model (\ref{eq:Schrod}) in the thermodynamic limit, the boundary
conditions at the two ends of the chain are not important. For
convenience we choose hard wall boundary conditions, $t_{0} =
t_{L} = 0$, $L$ being the length of the chain.  We can solve
Eq.\ (\ref{eq:Schrod}) recursively in terms of a sequence of hermitian $N
\times N$ matrices $a_n$,
\begin{eqnarray}
  a_n \psi_{n+1} &=& t_n^{\dagger} \psi_n,\ \
  a_{n} = - t_n^{\dagger} 
    \left(\varepsilon + a_{n-1} \right)^{-1}
    t_{n}^{\vphantom{\dagger}}. \label{eq:recurs}
\end{eqnarray}
The boundary condition at $n=0$ implies
$a_0 = 0$.
Evaluating the Schr\"odinger equation at $n=L$ then yields that
$\varepsilon$ is an eigenvalue of ${\cal H}$ if and only if
\begin{equation} \label{eq:eigval}
  \varepsilon \psi_{L} = - t_{L-1}^{\dagger} \psi_{L-1} = - a_{L-1} \psi_{L},
\end{equation}
i.e.\ if $a_{L-1}$ has an eigenvalue $-\varepsilon$, or alternatively,
in view of Eq.\ (\ref{eq:recurs}), if $a_{L}$ has a diverging
eigenvalue.

As a statistical model for the $t_n$ that contains all the relevant
symmetries, we parameterize the $t_n$ in terms of the generators
\begin{equation}
  t_n = e^{W_{n}},
\end{equation}
where $W_{n}$ is a real (complex) matrix for $\beta=1$ ($2$), and
choose the matrices $W_{n}$ from independent Gaussian distributions 
with mean and variance given by
\begin{eqnarray}
  \langle (W_{n})_{\mu \nu}^{\vphantom{*}}
          (W_{n})^{*}_{\rho \sigma} \rangle &=& \case{1}{2} {w^2 \beta}
  [\delta_{\mu \rho} \delta_{\nu \sigma} - \case{1-\eta}{N} \delta_{\mu \nu}
  \delta_{\rho \sigma}], \nonumber \\
  \langle (W_{n})_{\mu \nu} \rangle &=& \case{1}{2} (-1)^{n}
          \Delta \delta_{\mu \nu}.
\end{eqnarray}
Here $\eta$ governs the fluctuations of $\mbox{tr}\, W_n$
\cite{preparation} and $\Delta$ measures the staggering of the hopping
parameter. ($\Delta$ is the gap size that the staggering would induce
in the absence of disorder.) We assume that the disorder and staggering
are weak, and that the energy is small compared to the bandwidth ($w^2,
\Delta, \varepsilon \ll 1$).

The matrix $a_L$ has eigenvalues $\alpha_{\mu}$, that can be parameterized
as $\alpha_{\mu} \equiv \tan(\phi_{\mu}/2)$ ($\mu=1,\ldots,N$). As we
have discussed below Eq.\ (\ref{eq:eigval}), the energy $\varepsilon$ is 
an eigenvalue of the
Hamiltonian, if and only if there is an angle $\phi_{\mu}$ with
$\phi_{\mu} = \pi$. For general $\varepsilon$, however, none of the
$\phi_{\mu}$ will be equal to $\pi$. Nevertheless, we can use
the angles $\phi_{\mu}$ to compute the (disorder averaged) density of
states. Hereto we first note that $a_{L+2} = a_{L}$ in the absence of
disorder, staggering, and for $\varepsilon = 0$.  Then, taking
disorder, staggering, and a finite energy into account, and considering
the length $L$ as a fictitious ``time'', the angles $\phi_{\mu}$
perform a Brownian motion on the unit circle, which is such that upon
increasing $L$, they move around the circle in positive direction.
The rate at which the $\phi_{\mu}$ pass through $\pi$
as we increase $L$, i.e.\ their {\em current}, equals the number of
states per unit length $N(\varepsilon)$ with energy between $0$ and
$\varepsilon$. (This is a generalization of the node-counting theorem
\cite{Schmidt} used to compute the d.o.s.\ for $N=1$
\cite{EggarterRiedinger,OE}.) For comparison, we remark that, in the
absence of disorder, the angles $\phi_{\mu}$ move around at a constant
speed $\propto \varepsilon$, resulting in a constant d.o.s.
With disorder, their motion acquires a random (Brownian) component, which
dramatically affects their average speed, and hence the density of
states, as we shall see below.

\begin{figure}
\epsfxsize=0.7\hsize
\hspace{0.15\hsize}
\epsffile{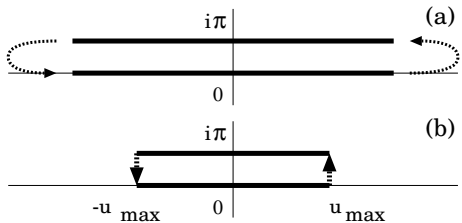}
\medskip

\caption{\label{fig:1}
(a) Two branches for the coordinate $u$. (b) In the simplified model,
the branches are truncated at $\pm u_{\rm max} = \ln(w^2/\varepsilon)$.}
\end{figure}

For a quantitative description a
different parameterization of the eigenvalues $\alpha_{\mu}$ proves
to be more convenient, 
\begin{equation}
  \alpha_{\mu} = \tan (\phi_{\mu}/2) = e^{u_{\mu}}.
\end{equation} 
The variables $u_{\mu}$ are restricted to the two branches
$\mbox{Im}\, u_{\mu} = 0$ and $\mbox{Im}\, u_{\mu} = \pi$ in the 
complex plane, see Fig.\ \ref{fig:1}a. We refer to these
as lower and upper branches, respectively. Noting that the $u_{\mu}$
are related to the angles $\phi_{\mu}$ on the unit circle, we 
see that a (fictitious) particle with
coordinate $u_{\mu}$ that vanishes on one of the branches at
$\pm\infty$ reappears at the opposite branch, as indicated by the
arrows in Fig.\ \ref{fig:1}a. Upon increasing $L$ by
two, the $u_{\mu}$ change according to
$u_{\mu} \to u_{\mu} + \delta u_{\mu}$, where, to lowest order in $w$, $\varepsilon$, and $\Delta$, the
average and variance of the increments $\delta u_{\mu}$ are
\begin{eqnarray}
  \langle \delta u_{\mu} \rangle &=&
    {2 \varepsilon} \cosh u_{\mu} + 2 \Delta
    + w^2 \beta \sum_{\nu \neq \mu} 
  \coth {u_{\mu} - u_{\nu} \over 2}, \nonumber \\
  \langle \delta u_{\mu} \delta u_{\nu} \rangle &=&
    4 w^2 \left[\delta_{\mu \nu} - (1-\eta)/N
    \right]. \label{eq:ubrown}
\end{eqnarray}
Taking $L$ as a continuous variable, their distribution function
$P(u_1,\ldots,u_N;L)$ obeys the Fokker-Planck equation \cite{VanKampen}
\begin{eqnarray}
  {\partial P \over \partial L} &=&
  w^2 \sum_{\mu, \nu} {\partial \over \partial u_{\mu}}
  \left( \delta_{\mu \nu} - {1-\eta \over N} \right) J {\partial
  \over \partial u_{\nu}} J^{-1} P \nonumber \\ &&
  \mbox{} - \sum_{\mu} {\partial \over \partial u_{\mu}}
  \left( {\varepsilon} \cosh u_{\mu} + \Delta \right) P,
  \nonumber \\
  J &=& \prod_{\mu < \nu} \sinh^{\beta}[(u_{\mu} - u_{\nu})/2].
  \label{eq:FP}
\end{eqnarray}

For large $L$ the solution of Eq.\ (\ref{eq:FP}) acquires a steady
state carrying a current $j(\varepsilon)$ equal to the integrated
d.o.s.\ $N(\varepsilon)$ in the thermodynamic limit.  Unfortunately,
except in the case $N=1$, where a solution in closed form is possible
\cite{OE}, it is notoriously problematic to find the steady-state
solution of a Fokker-Planck equation of the type (\ref{eq:FP}), due to
the lack of detailed balance \cite{VanKampen}.  Instead, below we
present a qualitative analysis of the Brownian motion process described
by Eq.\ (\ref{eq:FP}), illustrating the mechanisms that lead to the
anomalous behavior of the d.o.s.\ near $\varepsilon=0$. We give a
detailed account of the case $\eta = 1$, followed by a brief
discussion of the general case $\eta \neq 1$ and a comparison with
numerical simulations.

Let us first identify the relevant parameters in the Brownian motion
process (\ref{eq:FP}). There are $N$ fictitious
Brownian ``particles'' with
coordinates $u_{\mu}$ ($\mu=1,\ldots,N$) and diffusion coefficient $D =
w^2$. Three forces $F_{w}$, $F_{\Delta}$, and $F_{\varepsilon}$
act on the particles, arising from the presence of disorder,
staggering, and energy, respectively. The force $F_{w} = (w^2 \beta/2) \coth
[(u_{\mu} - u_{\nu})/2]$ is a repulsive interaction, with a hard-core on
the same branch and a soft core on different branches. (We adopt a
convention where the particles have unit mobility.) Staggering causes a
constant force field $F_{\Delta} = \Delta$ that favors motion to the
right on both branches. Finally, $F_{\varepsilon} = \pm \varepsilon
\cosh u$ pushes the particles to the left (right) on the lower (upper)
branch and thus causes the nonzero steady-state current. 
For small energies $\varepsilon \ll w^2$, motion is governed by the 
force $F_{\varepsilon}$ for large $|u|$, $|\mbox{Re}\, u| \gg u_{\rm max}$,
where
\begin{equation}
  u_{\rm max} = \ln(w^2/\varepsilon),
\end{equation}
while diffusion and the forces $F_{w}$ and $F_{\Delta}$ are dominant for 
small $u$, $|\mbox{Re}\,u| \ll u_{\rm max}$. Following
Ref.\ \onlinecite{EggarterRiedinger}, we now approximate our model by 
truncating the
branches at $| \mbox{Re}\, u| = u_{\rm max}$,
see Fig.\ \ref{fig:1}b, and adding a one-way move towards the upper
(lower) branches at the end-points, see Fig.\ \ref{fig:1}b. 
These
simplifications don't alter the functional dependencies of
$\rho(\varepsilon)$ for $\varepsilon \to 0$, though they may affect
prefactors.

\begin{figure}
\epsfxsize=0.94\hsize
\epsffile{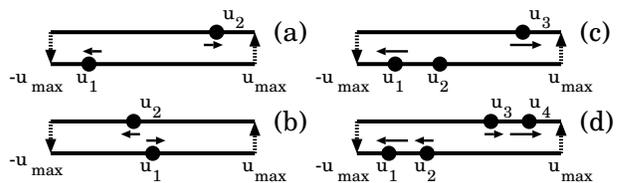}
\medskip

\caption{\label{fig:2}
Schematic picture of the forces on the fictitious particles in the
simplified model of Fig.\ \protect\ref{fig:1}b. For $N=2$, their mutual
repulsion slows the particles down and traps them near the end-points
(a).  When they eventually meet, the repulsion enhances travel (b), and
quickly restores the situation (a). The case $N=3$ is similar to $N=1$:
Two particles are trapped near the ends as a result of their repulsion,
while there is no net force on the third particle (c). For $N=4$, all
particles are trapped near the end points (d). The repulsive force on
the two middle particles $u_2$ and $u_3$ is smallest, and these two
dominate the current according to a scenario similar to the case $N=2$
(see a and b).}
\end{figure}

For $N=1$, the ``particle'' with coordinate $u$ needs a time $(2 u_{\rm
max})^2/2 D$ to diffuse to the end point of a branch, as one can verify
from a solution of the one-dimensional diffusion equation on the line
$-u_{\rm max} < u < u_{\rm max}$, with hard wall boundary conditions at
$u = - u_{\rm max}$ and absorbing wall boundary conditions at $u =
u_{\rm max}$ \cite{EggarterRiedinger}. Hence the current is
$j(\varepsilon) = N(\varepsilon) \propto {D / 4 u_{\rm max}^2}$, and
after differentiation with respect to $\varepsilon$, one finds 
Eq.\ (\ref{eq:N1div})
for the density of states. Diffusion effectively speeds up the
particle, explaining the enhancement of the d.o.s.\ relative to the
clean case \cite{Dyson1953,TheodorouCohen,EggarterRiedinger,OE}.  For
$N=2$, the picture is completely different.  As a result of their
mutual repulsion, the two particles with coordinates $u_1$ and $u_2$
get trapped near the end points, say at $u_{1} = - u_{\rm max}$ and
$u_{2} = u_{\rm max} + i \pi$. Now the particles have to diffuse out of
their traps against their repulsive interaction, until they eventually
meet at $\mbox{Re}\, u_1 = \mbox{Re}\, u_2$ and the repulsive force
$F_{w}$ starts to favor travel (see Figs.\ \ref{fig:2}a and b). Such a
process costs a large time, which can be calculated from the diffusion
equation for two particles on a line with hard wall boundary conditions
at $\mbox{Re}\, u_1 = - u_{\rm max}$ and $\mbox{Re}\, u_2 = u_{\rm
max}$ and absorbing boundary conditions at $\mbox{Re}\, u_1 =
\mbox{Re}\, u_2$. We find a current $j(\varepsilon) = (4 u_{\rm max}
F_{w}^3/D^2) \exp(-2 F_{w} u_{\rm max}/D)$, resulting in the
d.o.s.\ (\ref{eq:N2div}). (The prefactor $u_{\rm max}$ arises from the
degeneracy of the meeting point on the line.) We conclude that the
d.o.s.\ for two coupled chains has only a logarithmic singularity for
real hopping disorder, and a pseudogap for complex disorder. The strong
$\beta$-dependence of the d.o.s.\ stems from the $\beta$-dependence of
the interaction force $F_{w}$.

The qualitative behavior of the d.o.s.\ for general $N > 2$ depends
crucially on the parity of $N$ and closely resembles the scenarios we
have outlined above for $N=1$ and $N=2$.  If $N$ is even, all particles
get ``trapped'' near the ends of the branches, half of them on the
lower branch near $-u_{\rm max}$, and half of them on the upper branch
near $u_{\rm max}$, like in the case $N=2$, cf.\ Fig.\ \ref{fig:2}d. The
repulsive interaction force $F_{w}$ is smallest for the two particles
that are closest to the origin.  These two particles dominate the
current, resulting in a d.o.s.\ of the form (\ref{eq:two}). If $N$ is
odd, on the other hand, the picture is like that of the case $N=1$, see
Fig.\ \ref{fig:2}c.  All particles get trapped at the two ends, except
for one ``free'' particle, for which diffusion is not slowed down by
the interaction forces (the repulsive forces from the other particles
cancel exactly).  Hence, for odd $N$, the d.o.s.\ is of the form
(\ref{eq:N1}).  This even-odd effect is reminiscent of that found for
the conductance at $\varepsilon=0$, where the existence of a
delocalized state could be attributed to the existence of a similar
``free'' transmission eigenvalue \cite{BMSA}.

The effect of staggering $\Delta$ is to add a constant force pointed to
the right on both branches. As a function of the staggering strength,
the system alternates between behavior corresponding to even and odd
$N$. Repeating the above analysis, we find that with staggering, the
d.o.s.\ shows the maximum (\ref{eq:N1}) if $\Delta = (N+1-2j) w^2
\beta/2$, $j= 1,\ldots,N$, whereas it shows the minimum (\ref{eq:two})
if $\Delta = (N-2j) w^2 \beta/2$, $j=1, \ldots,N-1$. For all other
values of $\Delta$, for $\varepsilon \to 0$ we have
\begin{equation}
  \rho(\varepsilon) \propto \varepsilon^{\beta n-1},\ \ n =
  \min_{j=1,\ldots,N} \left|{2\Delta \over w^2 \beta} - N-1+2j\right|.
\end{equation}

The case of arbitrary $\eta \neq 1$ is not much different from the
case $\eta=1$ we considered above; up to prefactors the
$\varepsilon$-dependence of the d.o.s.\ is not changed. The only
exception is the case $\eta=0$, $N=2$, when the center of mass $u_1 +
u_2$ is pinned. As a result, the degeneracy giving rise to the
logarithm in Eq.\ (\ref{eq:two}) is lifted, and the logarithmic
prefactor vanishes.

\begin{figure}
\epsfxsize=0.72\hsize
\hspace{0.12\hsize}
\epsffile{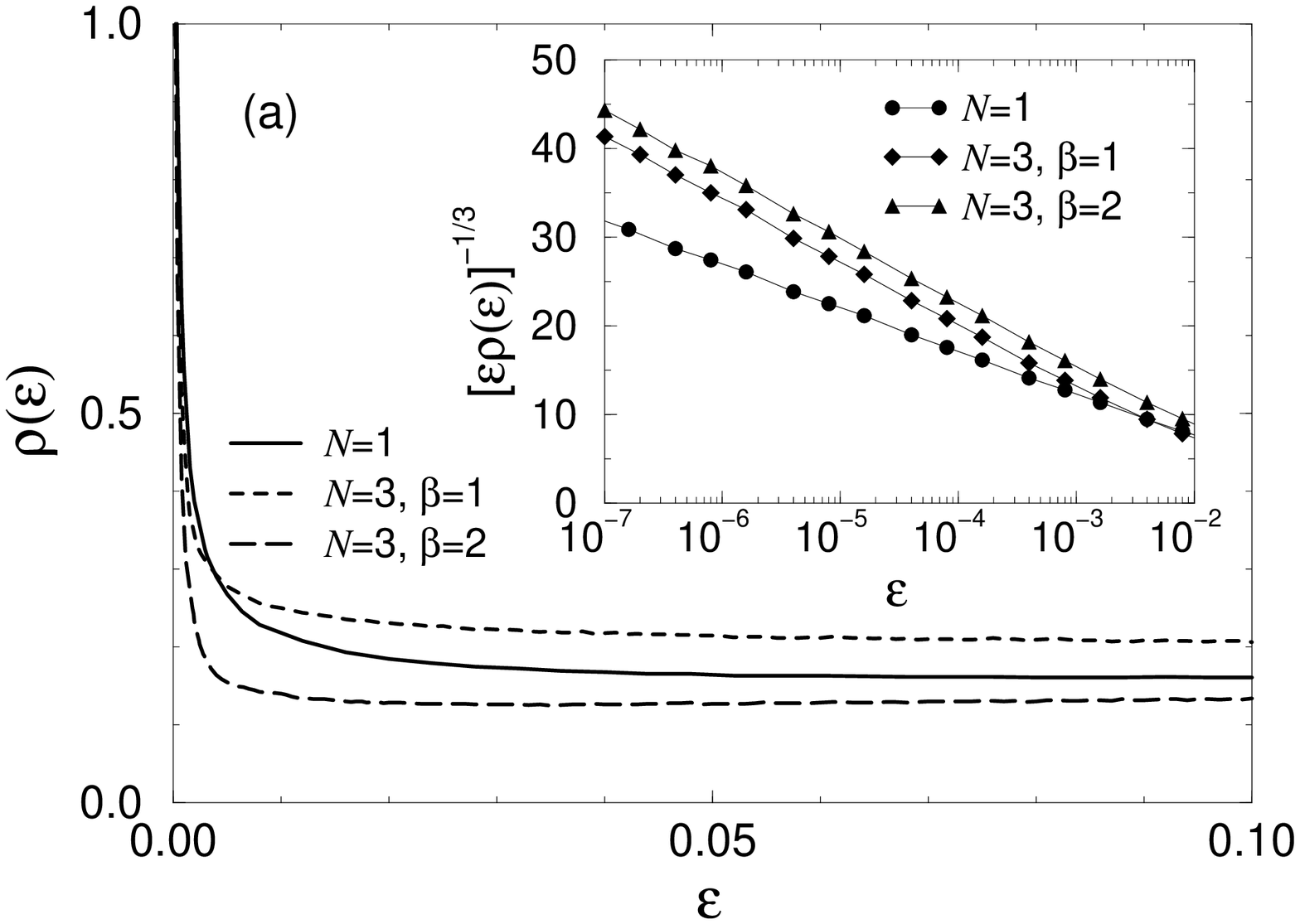}\vspace{-0.5cm}

\hspace{0.12\hsize}
\epsfxsize=0.72\hsize
\epsffile{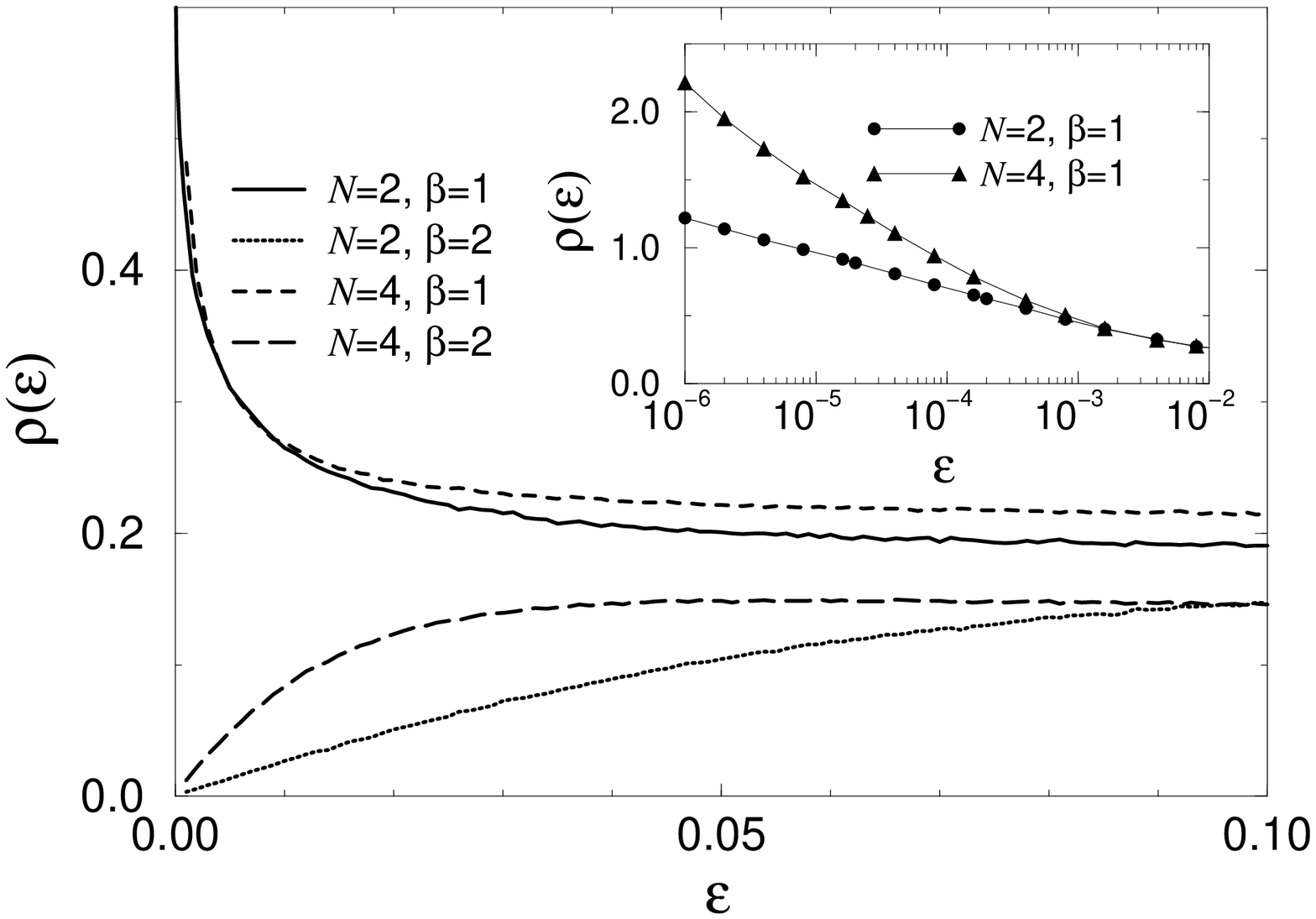}

\caption{\label{fig:3} Density of states, computed from numerical simulations for $N=1,3$ (a) 
and for $N=2,4$ (b).  The data shown on the linear scale are computed
for $L=200$ ($L=500$ for $N=1$) while the data in the insets are for
$L=10^5$ ($L=10^6$ for $N=1$).}
\end{figure}

We conclude with a comparison to numerical simulations for the
d.o.s. in a quantum wire on a square lattice with a width between
$N=1$ and $N=4$, and a length $L$ between $10^3$ and $10^5$. For 
$\beta=1$ and also for $N=1$ the hopping amplitudes are taken from a
uniform distribution in the interval [0.5,1.5], while for
$\beta=2$ the random flux model \cite{RandomFlux} is used, where
the randomness is introduced only via the random phases of the
hopping amplitudes. Results of an average over $4\times10^4$--$10^6$
disorder realizations are shown in Fig. \ref{fig:3}. The agreement with our theoretical results, Eqs.\ (\ref{eq:N1div}) and
(\ref{eq:N2div}), is excellent.

We would like to thank C.\ W.\ J.\ Beenakker, D.\ S.\ Fisher, and
B.\ I.\ Halperin for discussions.  PWB acknowledges support by the NSF
under grant nos.\ DMR 94-16910, DMR 96-30064, and DMR 97-14725.  CM
acknowledges a fellowship from the Swiss Nationalfonds.  The numerical
computations were performed at the Yukawa Institute Computer Facility.

\end{document}